\documentclass[aip,
amsmath,amssymb,
reprint,]{revtex4-1}

\usepackage{graphicx}\usepackage{dcolumn}\usepackage{bm}
\usepackage[utf8]{inputenc}
\usepackage[T1]{fontenc}
\usepackage{mathptmx}
\usepackage{etoolbox}

\usepackage{verbatim}
\usepackage[usenames,dvipsnames]{color}
\usepackage{bbm}
\usepackage{natbib}
\usepackage{xspace}
\usepackage{marginnote}
\usepackage{mathtools}
\usepackage{dsfont}
\usepackage{hyperref}
\usepackage{csquotes}

\usepackage{lmodern}

\usepackage{xcolor}
\usepackage{listings}

\usepackage{listings}
\usepackage{xcolor}

\usepackage{placeins}

\definecolor{basicColor}{HTML}{000000}
\definecolor{commentColor}{HTML}{008000}
\definecolor{keyWordColor}{HTML}{0000ff}
\definecolor{classColor}{HTML}{2681b0}
\definecolor{functionColor}{HTML}{795e26}
\definecolor{stringColor}{HTML}{a3151c}
\definecolor{parenthesisColor}{HTML}{319331}
\definecolor{codeBackground}{rgb}{0.99,0.99,0.99}

\lstdefinelanguage{cpp}{
	backgroundcolor=\color{codeBackground},  
	basicstyle=\footnotesize \ttfamily \color{basicColor} \bfseries,   
	breakatwhitespace=false,       
captionpos=b,                   
	commentstyle=\color{commentColor},   
	deletekeywords={...},          
escapechar=|,
	frame=single,                  
	language=C++,                
	keywordstyle=\color{keyWordColor},  
	morekeywords=[2]{std, cout, Real, RealVector, Matrix, AndersonDot, DoubleDot, RenormalizedPT, IteratedRG, Order, _1, _2, tf, Executor},
	keywordstyle=[2]\color{classColor},
	morekeywords=[3]{print, createModel, computeCurrent, computeCurrentKernel, computeMemoryKernel, computePropagator, diagonal, stationaryCurrent, stationaryState, begin, end, trace, transpose, Zero, LinSpaced, propagate, transform, size, real},
	keywordstyle=[3]\color{functionColor}, 
	identifierstyle=\color{black},
	stringstyle=\color{stringColor},      
	numbers=right,                 
	numbersep=5pt,                  
	numberstyle=\tiny\color{black}, 
	rulecolor=\color{black},        
	showspaces=false,               
	showstringspaces=false,        
	showtabs=false,                
	stepnumber=1,                   
	tabsize=5,                     
	title=\lstname,
	literate={\{}{{\textcolor{parenthesisColor}{\{}}}1  {\}}{{\textcolor{parenthesisColor}{\}}}}1  {)}{{\textcolor{parenthesisColor}{)}}}1   {(}{{\textcolor{parenthesisColor}{(}}}1    }

\robustify{\uparrow}
\robustify{\downarrow}
\robustify{\sum}
\robustify{\int}
\robustify{\nonumber}
\robustify{\cite}
\robustify{\footnote}

\renewrobustcmd{\Re}{{\text{Re}}}
\renewrobustcmd{\Im}{{\text{Im}}}

\newrobustcmd{\sdag}{{\bm{\dag}}}  \newrobustcmd{\tot}{\text{tot}} \newrobustcmd{\st}{\text{st}} 
\newrobustcmd{\deltah}{\bar{\delta}} \newrobustcmd{\sign}{\text{sgn}} \newcommand\order{\mathcal{O}}

\newrobustcmd{\K}{\mathcal{K}}
\ifdefined\L
\renewrobustcmd{\L}{\mathcal{L}}
\else
\newrobustcmd{\L}{\mathcal{L}}
\fi
\newrobustcmd{\Ord}{\mathcal{O}}

\newrobustcmd{\one}{\mathds{1}}   
\newrobustcmd{\ones}{\mathcal{I}}

\newrobustcmd{\ket}[1]{|#1\rangle}
\newrobustcmd{\bra}[1]{\langle#1|}
\newrobustcmd{\brkt}[1]{\left\langle #1 \right\rangle}
\newrobustcmd{\braket}[2]{\langle #1 | #2 \rangle}

\newrobustcmd{\Ket}[1]{\bm{|}#1\bm{)}}
\newrobustcmd{\Bra}[1]{\bm{(}#1\bm{|}}
\newrobustcmd{\Braket}[2]{\bm{(}#1\bm{|}#2\bm{)}}
\newrobustcmd{\Brkt}[1]{\bm{(} #1 \bm{)}}

\DeclareMathOperator{\Tr}{Tr}

\newrobustcmd{\Eq}[1]{Eq.~(\ref{#1})}
\newrobustcmd{\Eqs}[1]{Eqs.~(\ref{#1})}
\newrobustcmd{\eq}[1]{(\ref{#1})}
\newrobustcmd{\Fig}[1]{Fig.~\ref{#1}}
\newrobustcmd{\fig}[1]{\ref{#1}}
\newrobustcmd{\Figs}[1]{Figs.~\ref{#1}}
\newrobustcmd{\Sec}[1]{Sec.~\ref{#1}}
\newrobustcmd{\App}[1]{App.~\ref{#1}}
\newrobustcmd{\app}[1]{\ref{#1}}
\ifdefined\Ref
\renewrobustcmd{\Ref}[1]{Ref.~[\onlinecite{#1}]}
\else
\newrobustcmd{\Ref}[1]{Ref.~[\onlinecite{#1}]}
\fi
\newrobustcmd{\Refs}[1]{Refs.~[\onlinecite{#1}]}

\newrobustcmd{\Lst}[1]{Listing~\ref{#1}}

\renewrobustcmd{\App}[1]{\textbf{\textcolor{blue}{App.~\ref{#1}}}}
\renewrobustcmd{\app}[1]{\textbf{\textcolor{blue}{\ref{#1}}}}
\renewrobustcmd{\fig}[1]{\textbf{\textcolor{blue}{\ref{#1}}}}
\renewrobustcmd{\Fig}[1]{\textbf{\textcolor{blue}{Fig.~\ref{#1}}}}
\renewrobustcmd{\Figs}[1]{\textbf{\textcolor{blue}{Figs.~\ref{#1}}}}

\newrobustcmd{\InclDiagram}[1]{\raisebox{-0.3\height}{\includegraphics[height=0.45cm]{#1}}}

\makeatletter
\def\@email#1#2{\endgroup
	\patchcmd{\titleblock@produce}
	{\frontmatter@RRAPformat}
	{\frontmatter@RRAPformat{\produce@RRAP{*#1\href{mailto:#2}{#2}}}\frontmatter@RRAPformat}
	{}{}
}\makeatother

\newcommand{\cmt}[1]{}
\newcommand{\todo}[1]{}
\newcommand{\cut}[1]{}

\renewcommand{\cmt}[1]{\textcolor{magenta}{[#1]}}
\renewcommand{\todo}[1]{\textcolor{magenta}{TODO: #1}}

\begin{document}

\title{\texttt{RealTimeTransport}: An open-source C++ library for quantum transport simulations in the strong coupling regime}
\author{Konstantin Nestmann}
\email{konstantin.nestmann@ftf.lth.se}
\author{Martin Leijnse}\affiliation{
	Division of Solid State Physics and NanoLund, Lund University, Lund, Sweden
}
\author{Maarten R. Wegewijs}
\affiliation{
	Institute for Theory of Statistical Physics, RWTH Aachen, Aachen, Germany
}
\affiliation{
	JARA-FIT, Aachen, Germany
}
\affiliation{
	Peter Grünberg Institut, Forschungszentrum Jülich, Jülich, Germany
}

\date{\today}%
 
\begin{abstract}

The description of quantum transport in the strong system-reservoir coupling regime poses a significant theoretical and computational challenge that demands specialized tools for accurate analysis. \texttt{RealTimeTransport} is a new open-source C++ library that enables the computation of both stationary and transient transport observables for generic quantum systems connected to metallic reservoirs. It computes the Nakajima-Zwanzig memory kernels for both dynamics and transport in real-time going beyond traditional expansions in the bare system-reservoir couplings.
Currently, several methods are available:
(i) A renormalized perturbation theory in leading and next-to-leading order which avoids the low-temperature breakdown that limits the traditional theory.
(ii) Starting from this well-behaved reference solution a 2- and 3-loop self-consistent renormalization-group transformation of the memory kernels is implemented.
This allows refined quantitative predictions even in the presence of many body resonances, such as the Kondo enhancement of cotunneling.
This paper provides an overview of the theory, the architecture of \texttt{RealTimeTransport} and practical demonstrations of the currently implemented methods.
In particular, we analyze the stationary transport through a serial double quantum dot and showcase for the $T=0$ interacting Anderson model the complete time-development of single-electron tunneling (SET), cotunneling-assisted SET (CO-SET) and inelastic cotunneling resonances throughout the entire gate-bias stability diagram. We discuss the range of applicability of the implemented methods and benchmark them against other advanced approaches.

 \end{abstract}

\maketitle
\section{Introduction\label{sec:intro}}

The accurate description of quantum transport phenomena is fundamental to the understanding of quantum dot devices.
These are interesting platforms to study fundamental aspects of open quantum systems, featuring a rich interplay of non-equilibrium physics and many-body effects, as well as for various (quantum) technologies.~\cite{Cronenwett98, Sasaki00, Wiel2002, Su16, Gaudenzi17, Josefsson18}
Although their stationary transport properties are by now fairly well-understood, their dynamical aspects remain challenging.

From a methodological perspective an accurate description in extended parameter regimes is challenging.
Non-equilibrium Green's function methods are well-suited to study systems at low temperature, but struggle to treat strong Coulomb interactions which are typically incorporated perturbatively or in mean-field approximation.~\cite{Rammer09}
This has been improved by various means, in particular by applying renormalization group (RG) methods within this framework.~\cite{Metzner12}
On the other hand, a plethora of quantum master equation (QME) approaches exist,~\cite{Schoeller94, Gurvitz96, Koenig97, Timm08, Koller10} which treat the quantum dot energy scales non-perturbatively. These instead assume that the system-reservoir coupling is weak, which prohibits their usage at temperatures below the energy scale set by the tunnel rate.
Just like Green's function approaches, master equation-like methods can be improved by applying RG ideas, but in this case to approach stronger system-reservoir couplings and lower temperatures.~\cite{Schoeller09,Schoeller18}
The above mentioned methods are semi-analytical, which often aids the understanding of the results, and the work reported here follows this line.
Fully numerical approaches such as scattering state numerical renormalization groups,~\cite{Anders08, Nghiem14} (time-dependent) density matrix renormalization groups~\cite{Daley04, White04, Schollwoeck05} or quantum monte carlo methods~\cite{Rubtsov05, Muehlbacher08, Werner09, Cohen15}, just to name a few, have also been successfully applied.

In this paper we present \texttt{RealTimeTransport}, a C++  library which implements several methods
to analyze both transient and stationary quantum transport quantities based on the Nakajima-Zwanzig memory kernel of the open system.~\footnote{The library is available at \url{https://github.com/kn-code/RealTimeTransport}}
As the name suggests, the package is consistently based on the time-domain representation of the dynamics and transport. It thereby maintains a useful connection to the intuition of traditional quantum master equations (QMEs): inside the \enquote{black box} one can still identify rates for state transitions and Bloch coherence-vector dynamics.
At the same time it fixes shortcomings of these QMEs by fully accounting for renormalization and memory effects.
Key technical steps of our approach were heavily inspired by the so-called \enquote{$E$-flow} RG
formulated in frequency domain ($E$), which was successfully used to analytically and numerically describe the non-equilibrium Kondo effect in the stationary and time-dependent case.~\cite{Pletyukhov12, Bruch22}
We instead remain in the time domain noting that there the memory kernel is typically well-behaved
and can be efficiently resolved using well-developed Chebyshev interpolation techniques.~\cite{Trefethen2019}

Roughly speaking our approach works as follows: In ordinary bare perturbation theory one initially eliminates all dissipation from the problem by setting the coupling to the reservoir to zero. One then computes the corrections to this \enquote{too small dissipation} to increase it. In our approach we do exactly the opposite:
We initially analytically solve the problem in the physical limit of infinite temperature leading to \enquote{too large dissipation} and then compute corrections to this.
In prior work this was done by a \enquote{$T$-flow} from infinitely high temperature to the low temperature of interest.~\cite{Nestmann22} This involved solving a differential RG equation with respect to the physical temperature $T$, which at low $T$ approaches a fixed point (in the space of time-dependent memory kernels and vertex functions).

In the present work we perform this computation in a new way. We exploit that the \enquote{too large dissipation} computed at $T=\infty$ allows to construct an ansatz which is well-behaved even at $T=0$.
This reference solution incorporates all $T=\infty$ effects and is the leading term of a much simplified renormalized perturbation series.
In particular, this ansatz is very similar to the bare perturbation expression
whose virtual intermediate evolutions are regularized as $i0^+ \to \Sigma_{\infty}$, i.e.,
by replacing the problematic zero-dissipation by the $T=\infty$ Lindblad generator. Importantly, this superoperator-valued regularization is systematically derived from a physical limit and not put in \enquote{by hand}.
Our main result is that from this ansatz one can initiate a discrete RG iteration which converges stepwise to the same fixed point as the previous continuous $T$-flow, thus reducing the \enquote{too large dissipation} to the desired correct value.
Importantly, for the low temperature of interest this discrete iteration turns out to be more efficient then the continous $T$-flow.

Although the RG aspects of the outlined method make it stand out, it is firmly routed in the well-developed line of quantum master equation approaches for which various numerical packages have been published.
For example, the \texttt{QuTiP} (Quantum Toolbox in Python) package~\cite{Johansson12,Johansson13} implements phenomenological and leading order approaches in the system-reservoir coupling, such as the Lindblad, Bloch-Redfield and Floquet-Markov QMEs.
\texttt{QmeQ} (Quantum master equations for Quantum dot transport calculations)~\cite{Kirsanskas17} also implements several leading order methods, but additionally provides next-to-leading order approximations of the stationary state and particle/energy current through a consistent expansion of the memory kernel or the second-order von Neumann approach.

The paper is structured as follows: 
In \Sec{sec:methods} we first present the time-space perturbation theory for the memory kernel of the density operator and then develop the above mentioned renormalized version of this. Based on this we set up the new discrete iterative RG transformation and comment on some implementation details in \Sec{sec:library}.
In \Sec{sec:examples} we then showcase the practical library usage for several examples. In particular, we analyze the stationary transport through a serial double quantum dot and present the time evolution of the non-equilibrium interacting Anderson model in a magnetic field at zero temperature. This showcases the emergence of different strongly-correlated phenomena at distinct time-scales which is mapped out for all applied bias and gate voltages.
We note that the \emph{stationary} limit of this problem has by now been experimentally probed in detail in many types of quantum dots, ranging from semi-conductor heterostructures, carbon-based materials to molecular and atomic quantum dots.~\cite{Bockrath97, Nygard00, Franceschi01,Schleser05, Gaudenzi17}
Finally, we benchmark our methods against the density matrix renormalization group and quantum Monte Carlo approach.

\section{Computation of the memory kernel\label{sec:methods}}

In the following we set $\hbar=k_B=e=1$.

\subsection{Model and notation\label{sec:model}}

We consider a quantum dot system connected to several non-interacting electron reservoirs in the thermodynamic limit. Each of the reservoirs labeled by $r$ is assumed to be initially in a grand canonical equilibrium with temperature $T_r$ and chemical potential $\mu_r$. Thus the Hamiltonian describing the isolated reservoirs is given by
\begin{align}
	H_R = \sum_{r \nu} \int d\omega (\omega + \mu_r) a^\dagger_{r \nu}(\omega) a_{r \nu}(\omega),
\end{align}
and the initial state of the reservoirs is
\begin{align}
	\rho_R = \prod_r \frac{1}{Z_r} e^{-\beta_r \sum_\nu \int d\omega \omega a^\dagger_{r \nu}(\omega) a_{r \nu}(\omega)}.
\end{align}
Here $\nu$ denotes a generic channel index of the reservoirs, which is often just the spin $\sigma$. The field operators $a^\dagger_{r\nu}$ ($a_{r\nu}$) create (destroy) an electron with channel index $\nu$ in reservoir $r$ and anti-commute as
\begin{subequations}
	\begin{align}
		\big\{ a_{r \nu}(\omega), a^\dagger_{r' \nu'}(\omega') \big\} &= \delta_{r r'} \delta_{\nu \nu'} \delta(\omega-\omega'),
		\\
		\big\{ a_{r \nu}(\omega), a_{r' \nu'}(\omega') \big\} &= 0.
	\end{align}\label{eq:anticomm-a}\end{subequations}It is useful to introduce an additional particle-hole index $\eta$ and define
\begin{align}
	a_{\eta r \nu \omega} &\coloneqq \left\{\begin{array}{lrl}
		a^\dagger_{r \nu}(\omega) & \text{for} & \eta = +\\
		a_{r \nu}(\omega) & \text{for} & \eta = -
	\end{array}\right.
	.
\end{align}
For notational convenience we collect all indices into a multi-index written as a number,
\begin{align}
	1 \equiv \eta,r,\nu, \omega;\quad \bar{1} \equiv -\eta,r,\nu,\omega,
\end{align}
where the overbar indicates the inversion of the particle-hole index.
This allows us to rewrite the anti-commutation relations \eq{eq:anticomm-a} compactly as
\begin{align}
	\{a_1,a_2\}=\delta_{1 \bar{2}}.
\end{align}

Electrons can tunnel between the reservoirs and quantum dot system, which is captured by the tunneling Hamiltonian
\begin{align}
	H_T = \sum_{r \nu l} \int d\omega \sqrt{\lambda_{r\nu}} t_{r\nu l} d^\dagger_l a_{r \nu}(\omega) + \text{h.c.} \label{eq:H-T}
\end{align}
Here $d_l$ denotes a field operator of the quantum dot system and $l$ a (multi)index labeling the single particle states, including the spin and orbital quantum numbers. Importantly, the Fock space can always be constructed such that the field operators $d_l$ of the system and $a_{r \nu}(\omega)$ of the reservoir \textit{commute}, rather than anti-commute. This is explained in Sec. II. A. of~\Ref{Saptsov14}: one transforms from the conventional anti-commutation to commutation using the fermion parity operators of the system and reservoir, respectively.
This is crucial for tracing out the fermionic reservoir degrees of freedom later on, because it avoids keeping track of various minus signs.

We assume that the density of states $\lambda_{r\nu}$ and tunneling amplitudes $t_{r\nu l}$ are energy independent (wideband limit) and real-valued. They will later enter the transport equations via the spectral density, which we define as
\begin{align}
	\Gamma_{r l l'} \coloneqq 2\pi \sum_\nu \lambda_{r\nu} t_{r \nu l} t_{r \nu l'}.
	\label{eq:spectral-density}
\end{align}

The Hamiltonian $H$ describing the isolated quantum dot system is only constrained by the assumption that it should commute with the fermion parity (as any physical Hamiltonian should).~\cite{Wick52, Aharonov67} This is because the fermion parity superselection principle is explicitly incorporated into the superfermionic perturbation theory introduced later.
Thus, the total Hamiltonian we consider is given by
\begin{align}
	H_\tot = H + H_R + H_T.
\end{align}

\subsection{Superfermionic perturbation theory}

For initial total states that factorize, $\rho^0_\tot=\rho_0\otimes\rho_R$, the dynamics of the reduced density operator $\rho(t)\coloneqq\Tr_R \rho_\tot(t)$ can be described using the superoperator-valued propagator $\Pi(t)$ as $\rho(t)=\Pi(t) \rho_0$. The dynamics of $\Pi(t)$ follows the time-nonlocal Nakajima-Zwanzig quantum master equation,~\cite{Nakajima58,Zwanzig60}
\begin{align}
	\dot{\Pi}(t) = -i \L \Pi(t) - i \int_0^t ds \K(t-s) \Pi(s),
	\label{eq:qme-nonlocal}
\end{align}
where $\L\bullet = [H,\bullet]$ denotes the bare Liouvillian, $\K$ the memory kernel and the bullet an arbitrary operator argument. Our main goal is to compute the memory kernel $\K$. The perturbative expansion of $\K$ in the system-environment coupling is well-established,~\cite{Schoeller94,Koenig97,Leijnse08} but most commonly done in frequency space and in particular done at $\omega=0$, where the stationary state can be extracted.
However, in order to implement the renormalization group transformation presented in \Sec{sec:RG}, it is necessary to resolve the memory kernel either for all frequencies or alternatively for all times. Because the memory kernel as a function of time is typically well behaved and (at finite temperatures) exponentially decaying, it can be efficiently interpolated using Chebyshev polynomials,~\cite{Trefethen2019} which is done within \texttt{RealTimeTransport}. We will therefore 
outline the main steps to derive a time-space perturbation theory for $\K$ using the superfermionic formulation as presented in detail in \Refs{Saptsov12,Saptsov14}. This formulation is also at the heart of the more powerful methods presented in the subsequent sections.

The crucial starting point is the definition of superfield operators as
\begin{align}
	J^p_{1} \bullet \coloneqq \frac{1}{\sqrt{2}} \, \Big[ a_{1} \bullet + p (-\one)^{n_R} \bullet (-\one)^{n_R} a_{1} \Big]
	\label{eq:superfermion-def}
	,
\end{align}
where $(-\one)^{n_R}$ denotes the fermion parity of the reservoirs. These superfields are called \emph{superfermions}, because they act analogously to ordinary field operators $a_1$ in Liouville space. For example, they anticommute as
\begin{align}
	\left\{ J_{1}^{p_1}, J_{2}^{p_2} \right\} = \delta_{p_1 \bar{p}_2} \delta_{1 \bar{2}}
\end{align}
and obey a super-Pauli principle $(J^{p}_{1})^2=0$.~\cite{Saptsov14} Thus, $J^+_1$ ($J^-_1$) can be thought of as creating (annihilating) an excitation in Liouville space. 
By defining similar superfermions in the quantum dot systems Liouville space,
\begin{align}
	D^p_{1} \bullet \coloneqq \frac{1}{\sqrt{2}} \, \Big[ d_{1} \bullet + p (-\one)^n \bullet (-\one)^n d_{1} \Big]
	,
\end{align}
we can express the tunneling Liouvillian $\L_T\coloneqq [H_T,\bullet]$ as
\begin{align}
	\L_T = \sqrt{\lambda_1} t_1 D_1^p J_{\bar 1}^{\bar p}. \label{eq:L-T}
\end{align}
Here $\bar p \coloneqq -p$ and we implicitly sum/integrate over $p$ and $1=\eta, l, r, \nu, \omega$. Note that in writing \Eq{eq:L-T} we explicitly used the fermion parity superselection principle $(-\one)^{n_\tot}\rho_\tot(-\one)^{n_\tot}=\rho_\tot$, where $(-\one)^{n_\tot}$ denotes the fermion parity of the total system and $\rho_\tot$ any physical state.~\cite{Saptsov14}

Defining the unperturbed propagators of the dot and reservoir as $\Pi_0(t) \coloneqq e^{-i \L t}$ and $\Pi_R(t) \coloneqq e^{-i \L_R t}$ respectively, we can expand the full propagator $\Pi$ in orders of the tunneling Liouvillian $\L_T$. By implicitly summing over $n\in \mathbbm{N}_0$ and integrating over all intermediate times $\tau_i$ in a time-ordered fashion $t\geq\tau_1\geq\dots\geq\tau_n\geq0$ we obtain
\begin{align}
	\Pi(t) =& (-i)^n \big\langle \Pi_{0R}(t-\tau_1) \L_T \dots \L_T \Pi_{0R}(\tau_n) \big\rangle_R \label{eq:Pi-expand-1} \\
	=& (-i)^n \sqrt{\lambda_1} t_1 \dots \sqrt{\lambda_n} t_n \Pi_0(t-\tau_1) D_1^{p_1} \dots D_n^{p_n} \Pi_0(\tau_n) \notag \\
	&\times \big\langle \Pi_{R}(t-\tau_1) J_{\bar 1}^{\bar p_1} \Pi_R(\tau_1-\tau_2) \dots J_{\bar n}^{\bar p_n} \Pi_{R}(\tau_n) \big\rangle_R \label{eq:Pi-expand-2}
	.
\end{align}
In \eq{eq:Pi-expand-1} we used the shorthand notation $\Pi_{0R}=\Pi_{0}\Pi_R$ and $\langle \dots \rangle_R \coloneqq \Tr_R (\dots) \rho_R$. In \eq{eq:Pi-expand-2} we inserted \eq{eq:L-T} and used that by construction all system superoperators commute with those of the reservoir, allowing us to pull all of them to the left. The last line of \Eq{eq:Pi-expand-2} only involves reservoir fields and can be evaluated using a superoperator version of Wick's theorem,~\cite{Saptsov14} which decomposes the $n$-point expectation value $\langle J_1^{p_1} \dots J_n^{p_n} \rangle_R$ into sums of products of two-point expectation values  $\langle J_1^{p_1} J_2^{p_2} \rangle_R$.
A key simplification achieved by the superfermions \eq{eq:superfermion-def} is that out of the four two-point expectation values one obtains by setting $p_{1/2}=\pm$, only two are non-zero. These are given by
\begin{subequations}
	\begin{align}
		\langle J_1^{-} J_2^{+} \rangle_R &= \delta_{1 \bar{2}},
		\\
		\langle J_1^{-} J_2^{-} \rangle_R &= \delta_{1 \bar{2}} \tanh\left(\frac{\eta_2 \beta_{r_2} \omega_2}{2}\right).
	\end{align}\end{subequations}The internal $\omega_i$ integrations within \Eq{eq:Pi-expand-2} can then be incorporated into time-space contraction functions defined by
\begin{align}
	\gamma_{\eta_1 l_1,\eta_2 l_2}^p(\tau) \coloneqq \sqrt{\lambda_1 \lambda_2} t_1 t_2 \langle J_{\bar{1}}^{-} J_{\bar{2}}^{p} \rangle_R e^{-i \eta_1 (\omega_1 + \mu_{r_1}) \tau},
\end{align}where we again implicitly sum/integrate over all indices on the right hand side which do not occur on the left hand side. Evaluating the frequency integrals we obtain
\begin{subequations}
	\begin{align}
		\gamma_{\eta_1 l_1,\eta_2 l_2}^+(\tau) &= \delta_{\eta_1 \bar{\eta}_2} \tfrac{1}{2} \sum_r \Gamma_{r l_1 l_2} \bar\delta(\tau), \label{eq:contr-+} \\
		\gamma_{\eta_1 l_1,\eta_2 l_2}^-(\tau) &= \delta_{\eta_1 \bar{\eta}_2} \sum_r \frac{-i \Gamma_{r l_1 l_2} T_r}{\sinh(\pi T_r \tau)} e^{-i\eta_1 \mu_r \tau}. \label{eq:contr--}
	\end{align}\end{subequations}The $\bar\delta$ distribution occurring in $\gamma^+$ -- defined as $\int_0^t \bar\delta(t-s) f(s) ds = f(t)$ -- stems from the wideband limit taken in the definition of the tunneling Hamiltonian \eq{eq:H-T}.
Thus, we obtain the series expansion of the propagator in orders of the system-environment coupling, in a form where all reservoir degrees of freedom have been integrated out:
\begin{align}
	\Pi(t) =& (-i)^n \sum_{\text{contr.}} (-1)^{N_p} \big( \prod_\text{pairs $i<j$} \delta_{p_i +} \gamma_{i j}^{\bar{p}_j} (\tau_i - \tau_j) \big) \notag \\
	&\times \Pi_0(t-\tau_1) D_1^{p_1} \Pi_0(\tau_1-\tau_2) \dots D_n^{p_n} \Pi_0(\tau_n).
	\label{eq:Pi-PT}
\end{align}
Each term in \Eq{eq:Pi-PT} can be represented by a diagram, where vertices (superfermions), drawn as bullets,
\enquote{interrupt} the bare evolutions $\Pi_0$
and are contracted in pairs.
The memory kernel $\K$ is then given by the sum of all connected diagrams without \enquote{legs}, i.e., without the leftmost and rightmost bare propagator,
\begin{align}
	\K(t) =& (-i)^n \sum_{\text{connected}} (-1)^{N_p} \big( \prod_\text{pairs $i<j$} \delta_{p_i +} \gamma_{i j}^{\bar{p}_j} (\tau_i - \tau_j) \big) \notag \\
	&\times D_1^{p_1} \Pi_0(t-\tau_1)  D_2^{p_2} \dots \Pi_0(\tau_{n-2}) D_n^{p_n},
	\label{eq:K-bare-PT}
\end{align}
see the next section for examples. The factor $(-1)^{N_p}$ refers to the Wick-contraction sign, which is given by the number of crossings of contraction functions.

\subsection{Renormalized perturbation theory around the infinite-temperature limit}

The $\bar\delta$-function inside the memory kernel leads to a time-local contribution,
\begin{align}
	\K(t-s) = \Sigma_\infty \bar\delta(t-s) + \Sigma(t-s),
	\label{eq:decomposition-K}
\end{align}
where
\begin{align}
	\Sigma_{\infty} \coloneqq -\frac{i}{2} \sum_{\eta r l l'} \Gamma_{r l l'} D^+_{\eta l} D^-_{\bar\eta l'}.
\end{align}
Since the $\gamma^-$ contraction function \eq{eq:contr--} vanishes if all $T_r\rightarrow\infty$, we can infer that the Liouvillian
\begin{align}
	L_\infty \coloneqq L + \Sigma_{\infty}
\end{align}
generates the exact semigroup dynamics at infinite temperature:
\begin{align}
	\Pi_\infty(t) \coloneqq \lim_{T_r\rightarrow\infty} \Pi(t) = e^{-i L_\infty t}.
\end{align}
This infinite temperature solution can be used as a new reference in a \emph{renormalized} perturbation series.
The key technical step is to realize that two vertices connected by a $\gamma^+$ contraction can never have any other vertices between them, due to vanishing support of the intermediate time integrations, which enables the resummation of all $\gamma^+$ contractions into $\Pi_\infty$ propagators (see \Ref{Saptsov14} for details). It is now easy to obtain the rules of the renormalized perturbation theory, for which one has to adapt those of the bare one as follows: (i) only creation superfermions $D^+_1$ and $\gamma^-_1$ contractions are allowed and (ii) every bare Liouvillian $\L$ must be renormalized as $\L \rightarrow \L_\infty$, 
which includes replacing bare propagators as $\Pi_0 \rightarrow \Pi_\infty$.

\texttt{RealTimeTransport} implements the leading and next-to-leading order of this renormalized perturbation theory. In leading order, the renormalized memory kernel is explicitly given by
\begin{align}
	-i\Sigma^{(1)}(t) = \InclDiagram{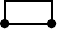} = - \gamma^-_{12}(t) D_1^+ \Pi_\infty(t) D_2^+.
	\label{eq:sigma-ren-1}
\end{align}
It is instructive to analyze the \enquote{worst case} $T\rightarrow 0$, where the bare perturbation theory suffers from a complete breakdown due to the very slow and oscillatory decay of the contraction function $\gamma^-(t) \propto e^{-i \eta_1 \mu_r t}/t$.~\cite{Nestmann21b} By contrast, the limit $T\rightarrow 0$ is unproblematic in \Eq{eq:sigma-ren-1} because the intermediate renormalized propagator is still exponentially decaying towards the infinte-temperature stationary state $\Pi_\infty(t)\rightarrow \tfrac{1}{d}\Ket{\one}\Bra{\one}$, which is a left zero eigenvector of the $D^+_2$ creation superfermion on the right.~\cite{Saptsov14}

The next-to-leading order corrections are
\begin{align}
	-i\Sigma^{(2)} = \InclDiagram{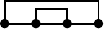} +
	\InclDiagram{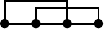},
\end{align}
where the diagrams are given by
\begin{align}
	\InclDiagram{diagrams/memory_kernel_2.pdf} =&  \gamma^{-}_{14}(t)  \gamma^{-}_{23}(\tau_1-\tau_2) D^+_{1} \Pi_\infty(t-\tau_1) D^+_{2} \notag \\
	&\times \Pi_\infty(\tau_1-\tau_2) D^{+}_{3} \Pi_\infty(\tau_2) D^{+}_{4}, \label{eq:sigma-ren-2} \\
	\InclDiagram{diagrams/memory_kernel_3.pdf} =& - \gamma^{-}_{13}(t-\tau_2)  \gamma^{-}_{24}(\tau_1) D^+_{1} \Pi_\infty(t-\tau_1) D^+_{2} \notag \\
	&\times \Pi_\infty(\tau_1-\tau_2) D^{+}_{3} \Pi_\infty(\tau_2) D^{+}_{4} \label{eq:sigma-ren-3}.
\end{align}
Here we again integrate in a time-ordered fashion over all intermediate times $\tau_i$, i.e., $t\geq\tau_1\geq\tau_2\geq0$, and sum over all occuring multi-indices.

Importantly, the singularities of $\gamma^-_{ij}(\tau)$ at $\tau=0$ within \Eqs{eq:sigma-ren-1}--\eq{eq:sigma-ren-3} never contribute due to the algebra of the superfermions, which was shown in detail in App. A of \Ref{Nestmann21b}. However, they need to be implemented carefully in order to not cause numerical problems.

Finally, we mention that for non-interacting systems it can be shown that the renormalized perturbation theory terminates due to the super-Pauli principle.~\cite{Saptsov14} Specifically, for these systems the renormalized perturbation theory of order $n_l$ is exact, where $n_l$ denotes the number of single particle states. For example, this means that for a single spinful level (without interactions) \Eqs{eq:sigma-ren-1}--\eq{eq:sigma-ren-3} give the exact memory kernel.

\subsection{Renormalization group transformation\label{sec:RG}}

We now follow \Ref{Nestmann22} to derive a renormalization group transformation with a fixed point given by the time-dependent memory kernel $\Sigma$ [\Eq{eq:decomposition-K}]. Firstly, in each diagram we resum all connected subblocks without uncontracted lines to full propagators $\Pi$, which we draw as double lines. The remaining diagrams are called irreducible:
\begin{align}
	-i \Sigma = \InclDiagram{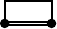} +
	\InclDiagram{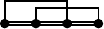} + \cdots .
	 \label{eq:Simga-diagrams}
\end{align}
For example, this leads to the diagram \eq{eq:sigma-ren-2} being already contained within the first diagram of \Eq{eq:Simga-diagrams}. Secondly, we define the effective 1-point vertex $\bar{D}_{1}$ as sum over all irreducible diagrams with 1 uncontracted line,
\begin{align}
	\bar{D}_{1} \equiv \InclDiagram{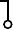} = \InclDiagram{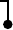} + \InclDiagram{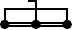} + \cdots, \label{eq:eff-one-point}
\end{align}
Because the effective vertex $\bar{D}_{1}$ represents a block of time, we need to distinguish between the time arguments of the leftmost vertex $t$, the uncontracted vertex $\tau$, and the rightmost vertex $s$, i.e., $\bar{D}_{1}\equiv \bar{D}_{1}(t,\tau,s)$. However, since we are only dealing with time-translation invariant Hamiltonians, the effective vertex only depends on time differences, $\bar{D}_{1} \equiv \bar{D}_{1}(t-\tau,\tau-s)$.

We can now utilize the effective vertex to express the memory kernel as~\cite{Nestmann22}
\begin{align}
	-i\Sigma(t) &= \InclDiagram{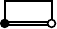} \label{eq:memory-kernel-exact-diagram}\\
	&= - \gamma^-_{12}(t-\tau_2) D_1^+ \Pi(t-\tau_1) \bar{D}_2(\tau_1-\tau_2,\tau_2).\label{eq:sigma-d-eff}
\end{align}
This result is formally exact and represents the memory kernel as a functional of the full propagator $\Pi$ and the effective vertex $\bar{D}_1$. However, since the propagator itself is determined by the memory kernel $\Sigma$ via \Eq{eq:qme-nonlocal}, we see that \Eq{eq:memory-kernel-exact-diagram} sets up a self-consistent RG transformation of the form
\begin{align}
	\Sigma = \mathcal{F} \big[ \Sigma, \bar{D}_1 \big]. \label{eq:RG-trafo}
\end{align}
Defining in an analogous way the effective two-point vertex as the sum over all irreducible diagrams with two free legs,
\begin{align}
	\bar{D}_{12} \equiv \InclDiagram{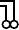} = \InclDiagram{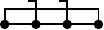}  + \cdots
	\label{eq:D2-first-term}
\end{align}
we find that the effective one-point vertex can be expressed as,~\cite{Nestmann22}
\begin{align}
	\InclDiagram{diagrams/effective_vertex_1.pdf} = \InclDiagram{diagrams/bare_vertex_1.pdf} +
	\InclDiagram{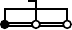} +
	\InclDiagram{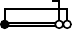} +
	\InclDiagram{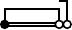},
	\label{eq:D1-eff-self-consistent}
\end{align}
which represents a self-consistent equation enslaved to \Eq{eq:sigma-d-eff} of the form
\begin{align}
	\bar{D}_1 = \mathcal{F}_1 \big[ \Sigma, \bar{D}_1, \bar{D}_{12} \big].
\end{align}
This indicates that the complete series \eq{eq:Simga-diagrams} can be rewritten as an infinite hierarchy of self-consistent equations for $n$-point effective vertices, where the $n$-point vertex is determined by the $n+1$-point vertex. This is similar in spirit to the flow equations of the functional renormalization group for Green's functions, where the self-energy is determined through an infinite hierarchy of $n$-particle vertices.~\cite{Metzner12}

In \Ref{Nestmann22}  \Eqs{eq:memory-kernel-exact-diagram}--\eq{eq:D1-eff-self-consistent} were approximately solved by transforming them into a continuous renormalization group flow using the environment temperature $T$ as flow parameter (dubbed \enquote{$T$-flow}). The main idea was to compute memory kernel corrections $\delta\Sigma$ which occur when the temperature is lowered in many small steps $\delta T$, thereby generating higher order coupling effects.

In \texttt{RealTimeTransport} we instead solve these equations iteratively in a 2- or 3-loop scheme as described below. We numerically checked that the solution generated this way is equivalent to the corresponding solution obtained from the continuous $T$-flow, but find that this discrete fixed-point iteration is more efficient than solving the differential $T$-flow RG equations.

In the 2-loop scheme we set the effective two point vertex $\bar{D}_{12}$ to zero and approximate the one-point vertex $\bar{D}_1$ via the first two terms of \Eq{eq:eff-one-point}. This gives a self-consistent scheme for the memory kernel of order $\order({D^+}^4)$.
Here we count the number of bare creation superfermions in each term, which are proportional to the number of contraction functions $\gamma^-$.
As initial approximation for the memory kernel we use the renormalized next-to-leading order memory kernel [\Eqs{eq:sigma-ren-1}--\eq{eq:sigma-ren-3}] and solve \Eq{eq:qme-nonlocal} to obtain the corresponding propagator.
We use these to obtain an initial approximation of $\bar{D}_1$, which we can then use to recompute the memory kernel and propagator. This scheme is iterated until the error between subsequent iterations is below some threshold.

In the 3-loop scheme we instead approximate the effective two point vertex $\bar{D}_{12}$ via the first term of \Eq{eq:D2-first-term} and expand the right hand side of \Eq{eq:D1-eff-self-consistent} such that all terms of order $\order({D^+}^5)$ are contained. This gives a self-consistent scheme for the memory kernel of order $\order({D^+}^6)$. The 3-loop iteration is instead started with the two-loop memory kernel. The idea is that this starting point should be \enquote{closer} to the three loop memory kernel, thus requiring fewer iterations to achieve convergence.

 \section{Library architecture\label{sec:library}}

\texttt{RealTimeTransport} currently implements the leading and next-to-leading order renormalized perturbation theory as well as the discussed 2- and 3-loop renormalization group transformations of the previous section.
Thus, the memory kernel has to be resolved in time-space.
Internally this is achieved through a Chebyshev interpolation with an adaptively chosen degree, based on a user-provided error goal. This approach leverages the fact that the memory kernel is typically a well-behaved, exponentially decaying function of time. For such smooth functions Chebyshev interpolations are nearly optimal in a well-defined sense~\cite{Trefethen2019} and vastly superior to interpolations on equidistant grids.
 
The primary challenge is that the occurring superoperators are large matrices of dimension $d^2 \times d^2$, where $d$ denotes the Hilbert-Fock space dimension, which need to be multiplied with each other many times when diagrams are evaluated by the numerical integration routines.
To reduce the size of matrices which need to be handled we use that the memory kernel has a block diagonal structure. Specifically, if an observable of the total system $N_{\text{tot}}=N+N_R$ is conserved and furthermore $[H,N]=[H_R,N_R]=0$, then
\begin{align}
	\Bra{s_1 s_2} \Sigma \Ket{s_{1'} s_{2'}} \neq 0 \, \Longrightarrow \, N_{s_1} - N_{s_2} = N_{s_{1'}} - N_{s_{2'}}
\end{align}
where $\Braket{A}{B}\coloneqq \Tr A^\dagger B$ denotes the Hilbert-Schmidt scalar product, $\Ket{s_1 s_2}=\ket{s_1}\bra{s_2}$ and $N\ket{s}=N_s\ket{s}$, see also App. F of \Ref{Saptsov12}. This means that each block of the memory kernel is characterized by a difference $\Delta N = N_{s_1} - N_{s_2}$, provided the basis vectors are correctly organized.
Each block of the memory kernel is thus internally represented by a matrix-valued Chebyshev interpolation object.

We furthermore use that the superfermions $D^+_1$ have a sparse structure inherited from the field operators $d_1$, i.e., almost all matrix elements are zero.
This allows us to separate each superfermion into a block structure defined by the memory kernel and only save the non-zero blocks as dense matrices.
The remaining linear algebra with dense matrices is handled using the well-established \texttt{Eigen} library.~\cite{EigenLib}

The memory kernel computation is implemented flexibly in a model-independent way. To create a new model one has to define a class that inherits from an abstract \texttt{Model} class and implement several methods, which provide, for example, the Hamiltonian, the field operators and information about the block structure of the memory kernel. Several important models have already been implemented in the library, such as the single and double quantum dot systems discussed in the next section.
 \section{Examples\label{sec:examples}}

In this section we exemplify the practical usage of the \texttt{RealTimeTransport} library, showcase each of the four implemented methods and discuss their scopes of applicability.

\subsection{Double quantum dot -- first order renormalized perturbation theory}

As a first example, we consider a spinless double quantum dot connected to two biased reservoirs. In the notation of \Sec{sec:model} the index $l\in\{1,2\}$ refers to the dot index and the reservoir index $\nu$ can be suppressed. Thus $d_l^\dagger$ ($d_l$) creates (annihilates) an electron in the dot $l$. We take the Hamiltonian of the double dot to be
\begin{align}
	H = \epsilon_1 n_1 + \epsilon_2 n_2 + U n_1 n_2 + \Omega (d_1^\dagger d_2 + d_2^\dagger d_1),
\end{align}
where $n_l\coloneqq d_l^\dagger d_l$ is the occupation of dot $l$ with energy $\epsilon_l$, $U$ the Coulomb interaction and $\Omega$ the hybridization between the dots. The coupling between each dot and its adjacent leads can be parameterized by $\Gamma_{r l} \coloneqq 2\pi \lambda_r |t_{r l}|^2$ such that $\Gamma_{rll'} = \sqrt{\Gamma_{r l} \Gamma_{r l'}}$ [\Eq{eq:spectral-density}]. In the following we focus on a serial setup with $\Gamma_{L 1} = \Gamma_{R 2} \equiv \Gamma$, $\Gamma_{L 2} = \Gamma_{R 1} = 0$ and furthermore set $\epsilon_1 = \epsilon_2 \equiv \epsilon$.

An interesting feature of this model is that coherences, described by non-diagonal elements of the density matrix,
do not vanish in the stationary state, for which we illustrate the computation using the first order renormalized perturbation theory in \Lst{lst:double-dot}.
The complete code for this can be found in the \texttt{examples} folder of the repository.\cite{Note1}
Since the double dot model is already implemented in the library, we only need to select parameters (line \ref{line:dd-one}-- \ref{line:dd-params}) and create the model using the \texttt{createModel} function (line \ref{line:dd-createModel}).
The cost of computing the memory kernel is mainly determined by the maximum time up to which it should be computed and an accuracy goal for the interpolation of the memory kernel in time space.
After the memory kernel is computed (line \ref{line:dd-computeK}) the stationary state can be directly accessed (line \ref{line:dd-rhoStat}).

\begin{lstlisting}[language=cpp,caption={Computation of the stationary state using the first order renormalized perturbation theory.},label={lst:double-dot}]
// Double dot parameters						|\label{line:dd-one}|
Real e     = -1;  // Dot energies
Real U     = 5;   // Coulomb interaction
Real Omega = 2;   // Hybridisation
Real V     = 0.5; // Bias voltage
RealVector T{{1, 1}}; // Temperatures leads
RealVector mu{{V/2, -V/2}}; // Chem. potentials

// Serial setup: lead L - dot 1 - dot 2 - lead R
RealVector Gamma1{{1, 0}}; // Dot 1 - lead L
RealVector Gamma2{{0, 1}}; // Dot 2 - lead R	|\label{line:dd-params}|

// Create model
auto model = createModel<DoubleDot>(			|\label{line:dd-createModel}|
    e, e, U, Omega, T, mu, Gamma1, Gamma2);

// Computation parameters
Real tMax    = 10;    // Maximum simulation time
Real errGoal = 1e-6;  // Interpolation error goal

// Memory kernel computation
auto K = computeMemoryKernel(model,				|\label{line:dd-computeK}|
    RenormalizedPT::Order::_1, tMax, errGoal);

// Compute stationary state and print it
Matrix rho = K.stationaryState();				|\label{line:dd-rhoStat}|
std::cout << "rho =\n" << rho << "\n";
\end{lstlisting}

To access the particle currents into the reservoirs one has to compute an additional current kernel, which is shown in \Lst{lst:double-dot-current}. After the current kernel is computed (line \ref{line:dd-computeK_I}) the stationary current can be directly accessed by passing the previously obtained stationary state as a parameter (line \ref{line:dd-Istat}).
\begin{lstlisting}[language=cpp,caption={Computation of the stationary current.},label={lst:double-dot-current}, firstnumber = last]
// Current kernel computation
int r    = 0; // Left lead
auto K_I = computeCurrentKernel(model, r,   |\label{line:dd-computeK_I}|
    RenormalizedPT::Order::_1, tMax, errGoal);

// Compute current and print it
Real I = K_I.stationaryCurrent(rho);        |\label{line:dd-Istat}|
std::cout << "I = " << I << "\n";
\end{lstlisting}

\begin{figure*}[ht]
	\centering
	\includegraphics[width=\linewidth]{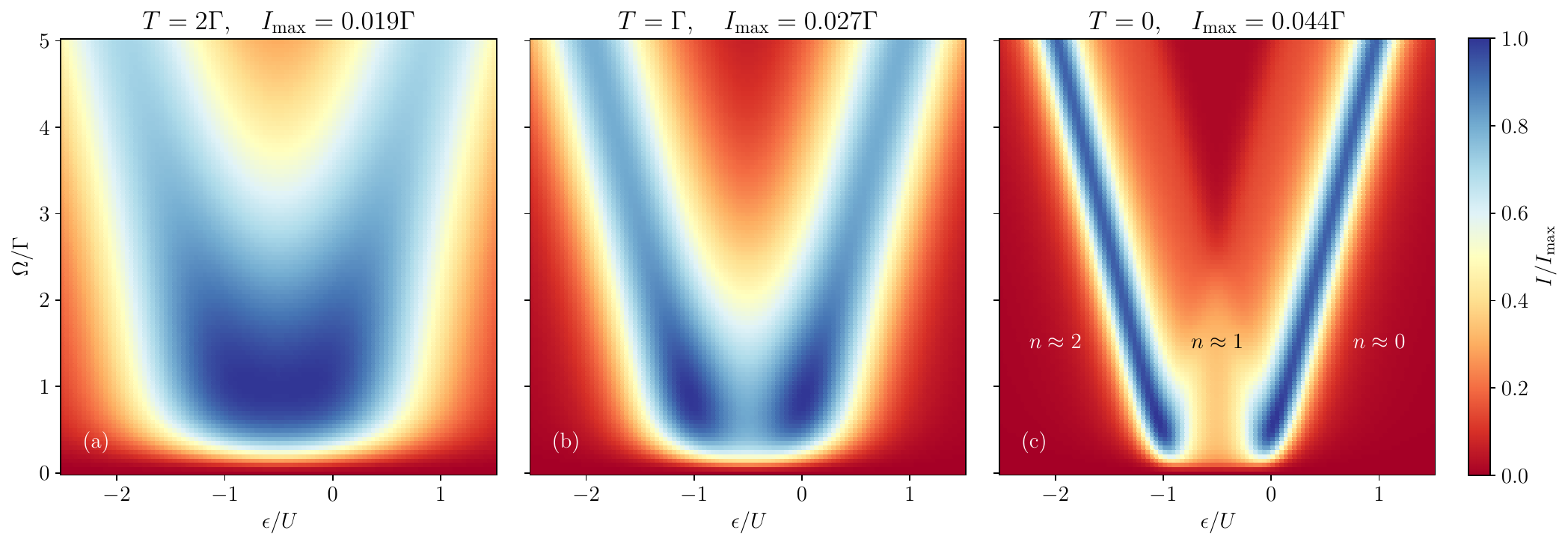}
	\caption{Leading order renormalized perturbation theory at $T=2\Gamma$, $\Gamma$ and $0$: Stationary current as function of interdot hybridization $\Omega$ and dot energies $\epsilon$ for a double quantum dot with interaction $U=5\Gamma$ at bias voltage $V=\Gamma/2$.}
	\label{fig:double-dot-current}
\end{figure*}
We illustrate the dependence of the stationary current on the hybridization $\Omega$ and dot energies $\epsilon$ by sweeping these parameters for three different temperatures $T\in\{0,1,2\}\Gamma$ in \Fig{fig:double-dot-current} and plot the ratio $I/I_\text{max}$, where $I_\text{max}$ denotes the maximum current within the scanned parameter range.
As expected one can clearly see that the broadening of the transport resonances is decreasing with temperature, but unlike in bare perturbation theory, the width remains finite for $T\rightarrow 0$.
Furthermore, as $\Omega \rightarrow 0$ the current vanishes because no tunneling between the dots is possible. This basic feature is completely missed by the popular Pauli master equation, which is even at high temperatures unable to describe the current correctly for $\Omega\lesssim\Gamma$ due to the role of coherences.

Other commonly used first order approaches, such as the von Neumann or Redfield approach, which do account for coherences break down when $T\lesssim\Gamma$, see \Ref{Kirsanskas17} for a detailed discussion.
The two blue lines of maximum current of width $V$ in \Fig{fig:double-dot-current}(c) are due to single electron tunneling (SET) and mark the transition of the stationary total dot occupation $n=n_1+n_2$ between $2\leftrightarrow 1$ and $1\leftrightarrow 0$ as indicated.
We note that cotunneling effects are important away from the resonances and not included consistently through the leading order renormalized perturbation theory. Here the next-to-leading order renormalized perturbation theory or the self-consistent RG methods should be utilized, which are illustrated in the next section.

\subsection{Transient Coulomb diamonds -- second order renormalized perturbation theory}

We now consider the single impurity Anderson model with energy $\epsilon$, magnetic field $B$ and Coulomb repulsion $U$ described by the Hamiltonian
\begin{align}
	H = \epsilon \left(n_\uparrow + n_\downarrow \right) + \tfrac{1}{2}B \left(n_\uparrow - n_\downarrow \right) + U n_\uparrow n_\downarrow.
\end{align}
The indices $l$ and $\nu$ of \Sec{sec:model} are spin indices $\sigma$. We take the tunnel rate $\Gamma$ to be spin- and reservoir-independent, i.e., $\Gamma_{r \sigma \sigma'}\equiv \Gamma \delta_{\sigma \sigma'}$ [\Eq{eq:spectral-density}].

The computation of transient occupations using the next-to-leading order renormalized perturbation theory is shown in \Lst{lst:Anderson-dot} and the complete code can also be found in the \texttt{examples} folder of the repository.~\cite{Note1}
After the model is set up similarly to before (line \ref{line:Ad-one}--\ref{line:Ad-createModel}) we use an additional \texttt{block} argument in the computation of the memory kernel and propagator (line \ref{line:Ad-computeK} and \ref{line:Ad-computePi}). This uses the fact that the memory kernel and propagator are block diagonal, since the particle number and spin component along the magnetic field are conserved. For the considered model the stationary state is encoded in a single block, which can also be used to compute transients as long as the initial state is described by a diagonal density matrix.
Thus, only computing a single block accelerates the computation. Omitting the \texttt{block} argument as in \Lst{lst:double-dot} defaults to computing the full memory kernel and propagator.
\begin{lstlisting}[language=cpp,caption={Computation of transient occupations $\langle n_\uparrow\rangle(t)$.},label={lst:Anderson-dot}]
// Create Anderson dot with two reservoirs      |\label{line:Ad-one}|
Real epsilon = -4;        // Dot energy
Real B       = -1;        // Magnetic field
Real U       = 10;        // Coulomb repulsion
RealVector T{{0, 0}};     // Temperatures leads
RealVector mu{{2, -2}};   // Chem. potentials
RealVector Gamma{{1, 1}}; // Tunnel rate to leads
auto model = createModel<AndersonDot>(         |\label{line:Ad-createModel}|
    epsilon, B, U, T, mu, Gamma);

// Computation parameters
Real tMax    = 5;    // Maximum simulation time
Real errGoal = 1e-4; // Interpolation error goal
int block    = 0;    // Only compute first block

// Compute memory kernel & propagator
auto method = RenormalizedPT::Order::_2;
auto K  = computeMemoryKernel(          |\label{line:Ad-computeK}|
    model, method, tMax, errGoal, block);
auto Pi = computePropagator(K, block);        |\label{line:Ad-computePi}|

// Set initial state: Basis 0, Up, Down, UpDown
Matrix rho0 = Matrix::Zero(4, 4);
rho0.diagonal() << 0, 1, 0, 0;

// Operator nUp
Matrix nUp = Matrix::Zero(4, 4);
nUp.diagonal() << 0, 1, 0, 1;

// Compute occupations Tr{nUp*rho(t)}
auto t = RealVector::LinSpaced(100, 0, tMax);
RealVector n(t.size());
for (int i = 0; i < t.size(); ++i)
    n[i] = (Pi(t[i], rho0) * nUp).trace().real();

// Print
std::cout << "t = " << t.transpose() << "\n"
          << "n = " << n.transpose() << "\n";
\end{lstlisting}
The computation of transient currents is shown in \Lst{lst:Anderson-dot-current}. The current kernel is computed (line \ref{line:Ad-computeK_I}) and used together with the previously obtained propagator in the function \texttt{computeCurrent} (line \ref{line:Ad-computeCurrent}), which returns an object containing a Chebyshev interpolation of the transient current. This allows to efficiently evaluate the current at arbitrary (vectors of) time arguments (line \ref{line:eval-current}).
\begin{lstlisting}[language=cpp,caption={Computation of the transient currents.},label={lst:Anderson-dot-current}, firstnumber = last]
// Compute current kernel
int r    = 0; // Left reservoir
auto K_I = computeCurrentKernel( |\label{line:Ad-computeK_I}|
    model, r, method, tMax, errGoal, block);

// Compute transient current and print
auto I         = computeCurrent(K_I, Pi, rho0);  |\label{line:Ad-computeCurrent}|
RealVector I_t = I(t);   |\label{line:eval-current}|
std::cout << "I = " << I_t.transpose() << "\n";
\end{lstlisting}
\begin{figure*}[t]
	\centering
	\includegraphics[width=\textwidth]{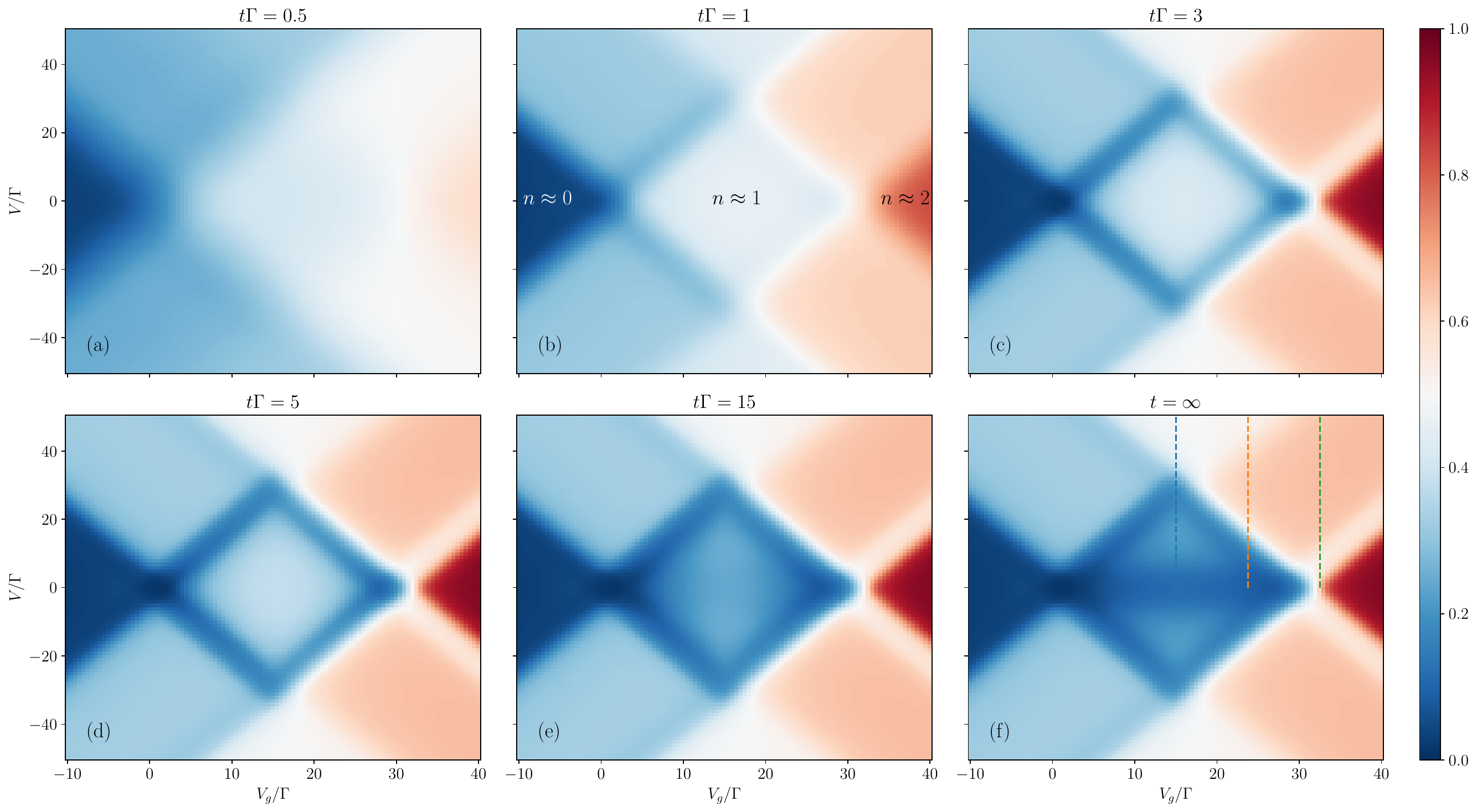}
	\caption{Next-to-leading order renormalized perturbation theory at $T=0$: Transient occupation of the excited state $\langle n_\uparrow\rangle(t)$ as function of bias voltage $V=2\mu_L=-2\mu_R$ and gate voltage $V_g=-\epsilon$ for an initially empty Anderson quantum dot $\rho_0=\ket{0}\!\bra{0}$ with strong interaction $U=30\Gamma$ and magnetic field $B=5\Gamma$ .}
	\label{fig:transient-occupation}
\end{figure*}
In \Fig{fig:transient-occupation} we show the transient occupation $\langle n_\uparrow\rangle(t)$ for an initially empty dot, $\rho_0=\ket{0}\!\bra{0}$, as function of the bias voltage $V$ and gate voltage $V_g=-\epsilon$ for several times $t$.
These results can be rationalized in detail based on well-established understanding of the stationary limit based on the transition rates.~\cite{Wegewijs01,Leijnse08}

In \Fig{fig:transient-occupation}(a)--(b)
fast SET processes fill the dot on a time scale $t\Gamma \sim 1$ to a total number of $ n\coloneqq n_{\uparrow}+n_{\downarrow}\approx 1$ and $2$ electrons as indicated in (b) (except in the $n\approx 0$ region where the dot is already stationary).
In these regions, the rate of filling is essentially independent
of the spin value, allowing the excited level $\ket{\!\!\uparrow}$ to be rapidly occupied with the same probability as the ground level $\ket{\!\!\downarrow}$.
In the $n\approx 1$ Coulomb diamond [white center in \Fig{fig:transient-occupation}(b)] this value $n_\uparrow\approx 1/2$ exceeds the stationary probability [\Fig{fig:transient-occupation}(f)] and multi-stage decay sets in.
At first this decay only occurs along a band of width $\propto B$ \emph{inside} the Coulomb diamond
[dark blue in \Fig{fig:transient-occupation}(c)].
Here the excess energy $B$ of the excited state $\ket{\!\!\uparrow}$ enables SET relaxation 
indirectly via the empty state $\ket{0}$ to the ground state $\ket{\!\!\downarrow}$ at rate $\propto \Gamma$.
We point out that this band is known as the cotunneling-assisted SET (CO-SET) regime in \emph{stationary} transport, since there the excited level can only be populated by inelastic cotunneling, see~\Refs{Leijnse08, Gaudenzi17} and references therein.
Here, on the other hand, this band is instead populated by fast SET into the empty dot due to our choice of initial state. Indeed, at this point in time ($t\Gamma \sim 1$) cotunneling has had no time yet to make an impact on the \emph{time-dependent} transport, as its rate scales as $\propto \Gamma^2/U$.
This impact is seen only in \Fig{fig:transient-occupation}(d)--(e)
where the excitation probability has been reduced uniformly throughout the $n\approx 1$ diamond where SET is energetically forbidden.
Here $\ket{\!\!\uparrow} \to \ket{\!\!\downarrow}$ relaxation occurs directly by
inelastic cotunneling processes involving \emph{both} reservoirs and releasing the excess energy $B$.
In \Fig{fig:transient-occupation}(f) this reduction has continued at low bias $|V| < B$
whereas it stalls at bias $|V| > B$ in the two visible triangles:
only in these regimes \emph{excitation} $\ket{\!\!\downarrow} \to \ket{\!\!\uparrow}$ by inelastic cotunneling is energetically possible while relaxation by CO-SET is forbidden.
This cotunneling re-excitation of the dot effectively counteracts the decay
and establishes a sizeable stationary value in competition with cotunneling relaxation.~\cite{Leijnse08}
Importantly, \texttt{RealTimeTransport} allows this rationalization to be further quantified by making use of the time-dependent memory-kernel matrix accessible internally.
For example, one interesting avenue would be to construct an effective time-dependent rate picture by converting this memory kernel to a time-local generator.~\cite{Megier20,Nestmann21a}

\subsection{2-loop RG corrections}

\begin{figure}[t!]
	\centering
	\includegraphics[width=\linewidth]{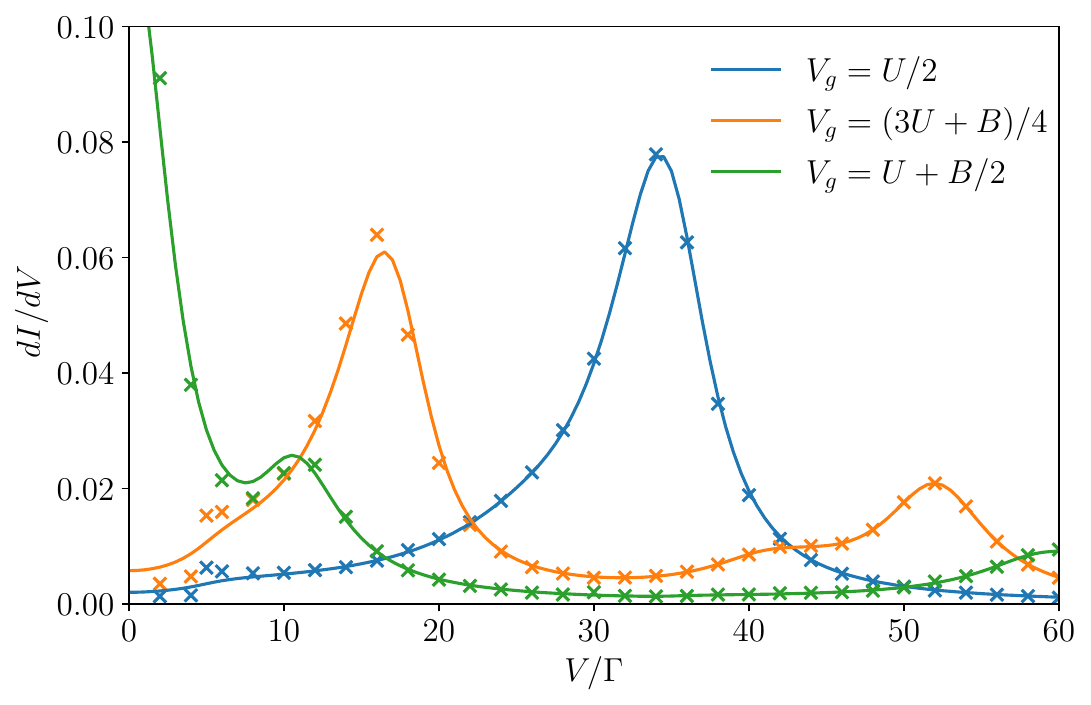}
	\caption{
		Nonlinear conductance $dI/dV$ at $T=0$ for three gate voltages $V_g=-\epsilon$ along the indicated cuts in \Fig{fig:transient-occupation}(f) as predicted by the next-to-leading order renormalized perturbation theory (full lines) and 2-loop renormalization group (crosses) for the same parameters as in \Fig{fig:transient-occupation}.
		\label{fig:method-comparison}
	}
\end{figure}

We now investigate when it is important to complement the renormalized perturbation theory with the self-consistency imposed by the 2-loop renormalization group transformation.
We plot in \Fig{fig:method-comparison} the (stationary) nonlinear conductance $dI/dV$ predicted by both methods across the lines indicated in \Fig{fig:transient-occupation}(f).
The conductance is computed using a central difference $dI/dV\approx (I(V+\delta V/2) - I(V-\delta V/2)) / \delta V$ with $\delta V=0.01\Gamma$.
Since we have a substantial magnetic field $B$ the Kondo resonance is not strongly developed here, even at the particle hole symmetric point $\epsilon=-U/2$ (blue curve).
Consistent with this, the RG corrections are moderate and occur where they are expected: namely inside the Coulomb diamond they suppress (enhance) the conductance below (above) the onset of ICT at $V \approx B=5\Gamma$ (blue curve). This has the effect of sharpening the ICT onset of the conductance relative to the renormalized perturbation theory.
Likewise, the SET signature of the same spin-flip resonance is sharpened.
This is in line with the general spirit of our approach as outlined in the introduction: We first introduce \enquote{too much} dissipation to subsequently compute corrections which reduce it in a non-trivial way.

From a user perspective, performing a calculation with the renormalization group methods is very similar to the perturbation theory methods and is shown in \Lst{lst:RG}. Because the computations are more time consuming it can be useful to parallelize them. We implemented multi-threading capabilities using the \texttt{Taskflow} library.~\cite{Huang22} This allows to define an \texttt{Executor} with a given number of threads (line \ref{line:executor-def}) that can be passed as an additional argument to compute the memory and current kernels (lines \ref{line:executor-K} and \ref{line:executor-Pi}).
\begin{lstlisting}[language=cpp,caption={Multi-threaded computation of the memory and current kernel using the RG  methods.},label={lst:RG}]
// Set up model and define parameter as before
// Compute memory kernel with 8 threads
tf::Executor executor(8);          |\label{line:executor-def}|
auto method = IteratedRG::Order::_2;
auto K  = computeMemoryKernel(model,    |\label{line:executor-K}|
    method, tMax, errGoal, executor);
auto Pi = computePropagator(K);

// Compute one block of the current kernel
auto K_I = computeCurrentKernel(K, Pi, r,    |\label{line:executor-Pi}|
    method, tMax, errGoal, executor, block);
\end{lstlisting}

\begin{figure*}[ht]
	\centering
	\includegraphics[width=0.9\linewidth]{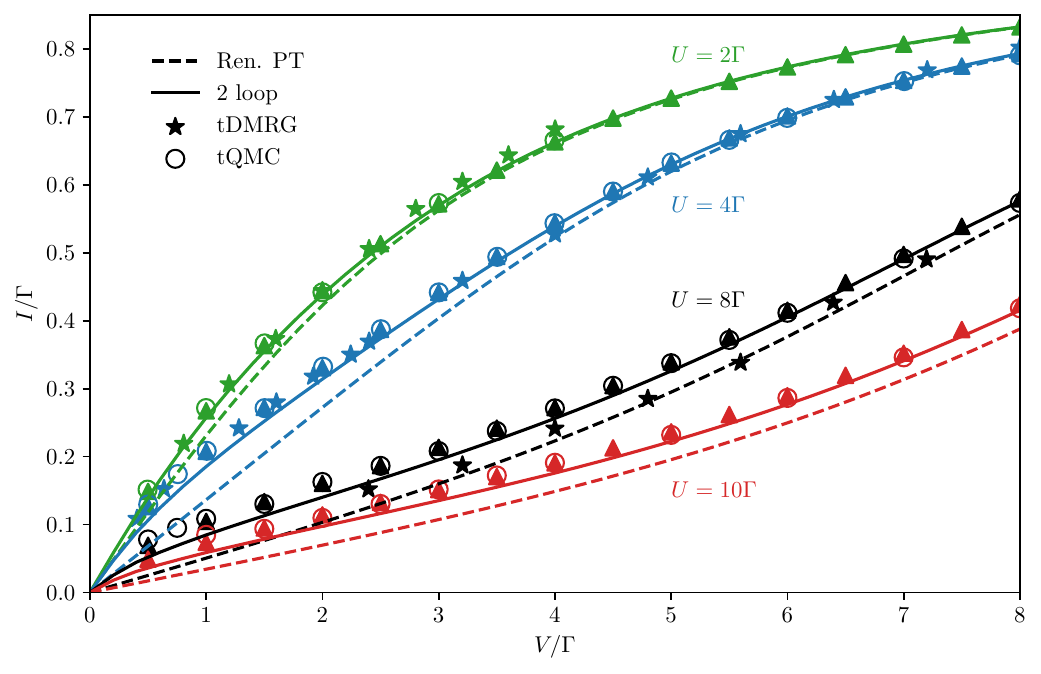}
	\caption{
		Comparison of the  stationary current at $T=0$ as function of the bias $V$ at the symmetry point $\epsilon=-U/2$ and magnetic field $B=0$:
		next-to-leading order renormalized perturbation theory, 2- and 3-loop RG, tDMRG and the tQMC method.
		\label{fig:method-comparison-2}
	}
\end{figure*}

\subsection{3-loop RG corrections}
Finally, we turn to the 3-loop renormalization group corrections at $T=0$.
We focus on the particle-hole symmetric point $\epsilon=-U/2$ and the most challenging case of zero magnetic field, $B=0$. In this case the Kondo effect becomes important for small bias $V$, which is challenging to describe for many theoretical methods.~\cite{Eckel10} In \Fig{fig:method-comparison-2} we compare the stationary currents obtained from our next-to-leading order renormalized perturbation theory and our 2- and 3-loop renormalization group with the time-dependent density matrix renormalization group (tDMRG) and the real-time quantum Monte Carlo method (tQMC).
The data for the latter two were taken from \Refs{Werner09, Meisner09, Werner10}, but note also \Ref{Eckel10} for further comparisons with the functional renormalization group and the iterative real-time path integral approach. For small interaction $U=2\Gamma$ we find that all methods are largely consistent with each other. The renormalized perturbation theory consistently predicts a current that is too small at low bias $V$, especially for larger interactions shown up to $U=10\Gamma$.
This can still be seen in our 2-loop RG, however to a much smaller extent. The 3-loop corrections only become important if $V$ is very small and $U$ is large. Even there we find good overall agreement between the currents predicted by our 3-loop RG and the tQMC method. We note that our discrete RG iteration reproduces the data from the continuous $T$-flow,~\cite{Nestmann22} confirming the expectation that they converge to the same fixed point.

\section{Summary\label{sec:summary}}

We have presented the open source C++ \texttt{RealTimeTransport} library which is available at \url{https://github.com/kn-code/RealTimeTransport}
to model stationary and transient quantum transport phenomena. The library currently implements four methods: the leading and next-to-leading order renormalized perturbation theory around the infinite temperature limit as well as more advanced 2- and 3-loop renormalization group transformations.
We exemplified the practical usage of each of these methods and compared their ranges of applicability.

In contrast to the ordinary bare perturbation theory in the tunnel coupling, our renormalized perturbation theory can also be applied all the way down to zero temperature. We have shown that it captures sequential and cotunneling effects in leading and next-to-leading order, respectively. However, especially at low temperatures even higher order effects become important. These can be systematically incorporated using our self-consistent perturbative renormalization group transformation of the memory kernel [\Eq{eq:RG-trafo}] of increasing loop orders.
For an Anderson dot we found that the most advanced 3-loop corrections contribute noticeably only if the system is close to the Kondo resonance. In this case we found overall good agreement between our 3-loop iteration and the tQMC method at finite bias.

The version of the library described in this paper is  \texttt{RealTimeTransport 1.0}.

\section*{Acknowledgments}

K.N. acknowledges funding from the European Union under the Horizon
Europe's Marie Sk{\l}odowska-Curie Project 101104590.
M.L. acknowledges funding from NanoLund, the Swedish Research Council (Grant Agreement No. 2020-03412) and the European Research Council (ERC) under the European Union's Horizon 2020 research and innovation programme under the Grant Agreement No. 856526.

\section*{Data Availability}

The data that support the findings of this study are available from the corresponding author upon reasonable request. 
 {}

\end{document}